\documentclass[12pt]{article}
\newcommand{\be}{\begin{equation}}
\newcommand{\ee}{\end{equation}}
\newcommand{\bea}{\begin{eqnarray}}
\newcommand{\eea}{\end{eqnarray}}
\newcommand{\bref}[1]{(\ref{#1})}
\newcommand{\nn}{\nonumber}
\def\str{{S}}
\begin{document}

\begin{center}
{\Large {\bf On the Cayley-Hamilton Theorem for Supermatrices}}

\vskip 10mm
{\large ~ Sotirios Bonanos$^1$ and Kiyoshi
Kamimura$^2$} \vskip 6mm
\medskip

${}^1$Institute of Nuclear Physics, NCSR Demokritos,\\
15310 Aghia Paraskevi, Attiki, Greece\\ and \\
${}^2$Department of Physics, Toho University Funabashi  
274-8510,  Japan \\
\vskip 6mm
 {\small e-mail:\
 {sbonano@inp.demokritos.gr\\
kamimura@ph.sci.toho-u.ac.jp} }\\

\medskip

\end{center}
\vskip 40mm
\begin{abstract}

We present a conjecture for expressing the coefficients in the Cayley-Hamilton theorem for supermatrices in terms of supertraces. The conjecture is tested for several supermatrix dimensions and unique results are obtained. Generating functions for determining these coefficients are also given.

\end{abstract}
\newpage

\section*{}
In the classical theory of matrices the Cayley-Hamilton theorem asserts that a given matrix $M$ satisfies its characteristic polynomial, the determinant of  the matrix $(x I-M)$. For an $n$-dimensional matrix this is a monic polynomial of degree $n$ in $x$, whose coefficients  can be expressed as polynomials in the traces of the first $n$ powers of the given  matrix.
 The generalization of the Cayley-Hamilton theorem to supermatrices is not straightforward because, when $M$ is a supermatrix, the superdeterminant of $(x I-M)$ is a {\it ratio} of polynomials in the variable $x$. 

Let the (even) supermatrix  $M$ have the block form 
\be
M= \left( \matrix{A & B\cr
C & D\cr}
\right),
\ee
where $A,\,B,\,C,\,D$ are, respectively, $p \times p,\,p  \times q,\,q  \times p,\,q  \times q$ dimensional matrices with $p+q=n$. The submatrices $A,\,D$ have even and $B,\,C$ have odd parity.
We will denote by $a(x)\,(d(x))$ the characteristic polynomial of the submatrix $A\,(D)$. Then the {\it characteristic  function} (the superdeterminant of $(x I-M)$) can be written as the ratio of  two 
polynomials in $x$ in two equivalent ways \cite{Kobayashi90, Morales94}:
\bea
h(x)&=&sdet(x I-M) \nn\\
&=&\frac{det[d(x)(x I-A)-B\, {\rm adj}(x I-D)C]}{d(x)^{p+1}}\label{hx1}\\
&=&\frac{a(x)^{q+1}}{det[a(x)(x I-D)-C\, {\rm adj}(x I-A)B]},\label{hx2}
\eea
where ${\rm adj}(M)$ denotes the adjoint (=determinant times inverse) of $M$. 

In \cite{Morales94}, the polynomial ${\cal P}(x)=a(x)^{q+1}d(x)^{p+1}$ has been proposed as the characteristic polynomial, which is of degree $2pq+p+q$ in the variable $x$. When $a(x)$ and $d(x)$ have no common factors,  it is shown in \cite{Kobayashi90}
 that $h(x)$ can be written  as $(a(x)+r)/(d(x)+t)$ where $r,\,t$ are polynomials with zero body and degree less than $a(x),\,d(x)$, respectively.  In this case the characteristic polynomial becomes  $P(x)=(a(x)+r)(d(x)+t)$, which is of degree $p+q$, i.e., the dimension of the supermatrix, just as for ordinary matrices.

 The definition $h(x)=sdet(x I-M)$ is invariant under general similarity transformations  $M \to U_M\,M\,U_M^{-1}$. When the similarity transformation supermatrix $U_M$ is block-diagonal, the polynomials appearing in the numerator and denominator of $h(x)$ are separately 
 invariant.\footnote{These statements follow from the definition of these polynomials in terms of determinants and the properties of determinants: $det(x I-UAU^{-1})=det(U(x I-A)U^{-1})=det(x I-A)$, which are true also for superdeterminants.}  This implies that the coefficients of $x$ in these polynomials are  invariants of the supermatrix under similarity transformations with a block-diagonal supermatrix $U_M$. 
One expects that, as for ordinary matrices, these coefficients will be expressible in terms of the supertraces of powers of $M$: $\str_j\equiv str(M^j)$. If $M$ satisfies identically a characteristic polynomial of degree $n_{max}$, then the supertrace of this identity implies that the supertraces of the first $n_{max}$ powers of $M$ must appear in the coefficients. 
 Higher powers of $M$ and corresponding supertraces will then be expressible in terms of these using the polynomial identity. But the supertraces are polynomials in the eigenvalues of $M$ whose number equals the dimension of $M$, so $n_{max}$ must be equal to $p+q$.

Observe that the coefficients of $x$ in the polynomials of the numerator and denominator of $h(x)$ are themselves polynomials in the matrix elements (only determinants and cofactors appear -- no inverses); and recall that the characteristic polynomial ${\cal P}(x)=a(x)^{q+1}d(x)^{p+1}$  is of degree $2pq+p+q$, while the reduced polynomial $P(x)=(a(x)+r)(d(x)+t)$ of degree $p+q$ is obtained when $a(x)$ and $d(x)$ have no common factors.
These considerations lead to the following conjecture:
\vskip 3mm

{\it The Cayley-Hamilton Theorem for supermatrices:}  A ($p,\,q$) supermatrix $M$,  whose diagonal submatrices ($A,\,D$) have eigenvalues ($a_i,\,d_i$) with distinct bodies ($\alpha_i\neq\delta_j$), satisfies the equation 
\be
a_{2pq}M^{n}+a_{2pq+1}M^{n-1}+....+a_{2pq+n}I=0, \label{superCH}
\ee
where $n=p+q$ and the $a_{k}$ are {\it polynomials} in the supertraces $\str_1,...,\str_n$  of $M$, which are homogeneous  of total  degree $k$ in the  elements of the matrix $M$, 
so that every term in the sum is of degree $2pq+p+q$ in the matrix elements.
\vskip 3mm

Using {\it Mathematica}, we have tested this conjecture for all values of $(p,\,q)$ satisfying $p+q<6$, as well as $(5,1)$ and $(1,5)$.
In all cases, the supertrace polynomials $a_{k}(\str_i)$ are determined {\it uniquely} (up to an overall numerical factor) by the assumptions stated. In fact, these polynomials  can be determined using a {\it diagonal} ($p,\,q$) supermatrix;  
and subsequently it can be verified that any such ($p,\,q$) supermatrix satisfies \bref{superCH} with the same expressions for $a_{k}$ as functions of the $\str_i$. In all cases examined the coefficient $a_{2pq}$ of the highest power of $M$ is the square of a polynomial of order $pq$ in the matrix elements, while the coefficient $a_{2pq+1}$ is the product of this polynomial with a second one (of order  $pq+1$). We give below the complete expressions of the super-Cayley-Hamilton identities for the lowest supermatrix dimensions ($p+q<5$) choosing the coefficient of  ${\str_1}^{2pq}\,M^{p+q}$ to equal unity:
\bea
(1,\,1)&:& {\str_1}^2\,M^2-\str_1\, \str_2\, M +\frac{1}{4} \left(-{\str_1}^4+ {\str_2}^2\right)I=0, \label{CH11}
\\
(2,\,1)&:&\left({\str_1}^2-\str_2\right)^2 M^3-\frac{1}{3}
   \left({\str_1}^2-\str_2\right) \left({\str_1}^3+3\, \str_1\,\str_2
   -4 \,\str_3\right) M^2\nn\\
   &-&\frac{1}{18} \left({\str_1}^6-9\,
   {\str_1}^4\str_2 +16\, {\str_1}^3\str_3 -9\, {\str_1}^2 {\str_2}^2
   +9\, {\str_2}^3-8\, {\str_3}^2\right) M\nn\\
   &+&\frac{1}{18}
\left({\str_1}^4-4\, {\str_1}\, {\str_3}+3\, {\str_2}^2\right)
\left({\str_1}^3-3 \,{\str_1}\, {\str_2}+2\, {\str_3}\right)\,I=0, \label{CH21}\\
 (1,\,2)&:&\left({\str_1}^2+\str_2\right)^2 M^3+\frac{1}{3}
   \left({\str_1}^2+\str_2\right) \left({\str_1}^3-3 \, \str_1\,\str_2
 -4 \,\str_3\right) M^2\nn\\
   &-&\frac{1}{18} \left({\str_1}^6+9\,
  {\str_1}^4 \str_2 +16\, {\str_1}^3 \str_3 -9 \,{\str_1}^2 {\str_2}^2
  -9 \,{\str_2}^3-8\, {\str_3}^2\right) M\nn\\
   &-&\frac{1}{18} \left({\str_1}^4-4\, {\str_1}\, {\str_3}+3\, {\str_2}^2\right)
\left({\str_1}^3+3 \,{\str_1}\, {\str_2}+2\, {\str_3}\right)\,I=0,\label{CH12}\\ 
(3,\,1)&:&  \left(
      {\str_1}^3 - 3\,{\str_1}\, {\str_2} + 
          2\, {\str_3}\right)^2 M^4  \nn\\&-& \frac{1}{2} \left( {\str_1}^3 - 3\, {\str_1}\,{\str_2}  + \
2 \,{\str_3}\right)\left( {\str_1}^4 - 4\,{\str_1}\, {\str_3}- 
      3\, {\str_2}^2 + 6\, {\str_4}\right)M^3  \nn\\&+& \frac{1}{16}[{\str_1}^8 + 4 \,{\str_1}^6{\str_2}  - 
          32 \,{\str_1}^5{\str_3}  - 6 \,{\str_1}^4 \left( {\str_2}^2 - 6\, {\str_4}\right)   \nn\\&+& 64\, {\str_1}^3{\str_2}\, {\str_3}  - 4\, {\str_1}^2\left(
        9 \,{\str_2}^3 + 18\, {\str_2}\, {\str_4} + 8\, {\str_3}^2\right)   \nn\\&+& 96\, {\str_1} \, {\str_2}^2 {\str_3}+ 9\, {\str_2}^4 - \
32 \,{\str_2}\, {\str_3}^2 + 36\, {\str_4}^2 - 36\, {\str_2}^2  {\str_4}] M^2 \nn\\&+& \frac{1}{48}[
         {\str_1}^9 - 12 \,{\str_1}^7 {\str_2}  + 24\, {\str_1}^6{\str_3}  + 18\, {\str_1}^5\left( {\str_2}^2 - 2\, {\str_4})\right)   \nn \\&+& 24 \,{\str_1}^4{\str_2}\, {\str_3}  - 12\, {\str_1}^3\left(3\, {\str_2}^3 - 
        6 \,{\str_2}\, {\str_4} + 8\, {\str_3}^2\right) \label{CH31} \\&-& 72\, {\str_1}^2{\str_3}\left( {\str_2}^2 -    2 \,{\str_4}\right)  -  8\, {\str_3}\left(9\, {\str_2}^3 - 18 \,{\str_2}\, {\str_4}  + 8 \,{\str_3}^2\right)  \nn\\&+& 3\,{\str_1}\left(27\, {\str_2}^4 - 36\, {\str_2}^2 {\str_4}  + 32\, {\str_2}\, {\str_3}^2- 36\, {\str_4}^2\right) ] M \nn\\
        &-& \frac{1}{96}\left({\str_1}^4-6\, {\str_1}^2 {\str_2} +3\, {\str_2}^2 +8\, {\str_1}\, {\str_3}-6\, {\str_4}\right) \nn \\
&\times&[{\str_1}^6-3\, {\str_1}^4 {\str_2} +9\, {\str_1}^2{\str_2}^2+9\, {\str_2}^3 -24\, {\str_1}\,  {\str_2}\, {\str_3} \nn \\
&-&8\, {\str_1}^3 {\str_3}+16\, {\str_3}^2+18\, {\str_4}( {\str_1}^2- {\str_2})]\,I=0, \nn\\
(2,\,2)&:&\left(
      {\str_1}^4 - 4 \,{\str_1}\,{\str_3} + 3\,{\str_2}^2\right)^2 M^4 \nn  \\&-& 2\left(
      {\str_1}^4 -4 \,{\str_1}\,{\str_3} + 3\,{\str_2}^2\right)\left(
          {\str_1}^3 {\str_2}- 3 \,{\str_1}\,{\str_4} + 2\,{\str_2}\,{\str_3}\right)M^3
  \nn  \\&+&
  [-\frac{1}{12}
             {\str_1}^{10} + \frac{3}{2} {\str_1}^6 {\str_2}^2 + \
{\str_1}^4\left(4\, {\str_3}^2 - 9 \,{\str_2}\, {\str_4}\right)     - \frac{32}{3}{\str_1}\,{\str_3}^3
\nn  \\&+& \frac{9}{4} {\str_1}^2\left( {\str_2}^4 + 4 \,{\str_4}^2\right)
               + 3\,{\str_2}^2\left(4\,{\str_3}^2 - 3\,{\str_2}\,{\str_4}\right)] M^2
 \nn  \\&+&[\frac{1}{12}{\str_1}^9{\str_2} - {\str_1}^7{\str_4} 
 +    {\str_1} \left(8\,{\str_3}^2{\str_4}-\frac{9}{4}{\str_2}^5 - 
              9\,{\str_2}\,{\str_4}^2\right)
  \label{CH22}
         \\&+&  2\,{\str_1}^6{\str_2}\,{\str_3}
               + 2\,{\str_1}^4{\str_3}\,
           {\str_4}+{\str_1}^3 {\str_2}
        \left(3\,{\str_2}\,{\str_4} - 8\,{\str_3}^2\right)   \nn  \\&+& 6 \,{\str_1}^2{\str_2}^3{\str_3}
             - \frac{3}{2}{\str_1}^5{\str_2}^3 + \frac{2}{3}{\str_2}
        {\str_3}\left(9\,{\str_2}{\str_4} - 8\,{\str_3}^2\right)]M   
 \nn  \\&+&
   \frac{1}{144}
 [\left({\str_1}^6+9\, {\str_1}^2 {\str_2}^2-8\, {\str_1}^3 {\str_3}+16\, {\str_3}^2-18\, {\str_2}\, {\str_4}\right)^2 \nn \\
&-&
9\left({\str_1}^4{\str_2}-3\, {\str_2}^3+8\, {\str_1}\, {\str_2}\,{\str_3}-6\, {\str_1}^2 {\str_4}\right)^2]\,I=0,   \nn     
    \eea  
We observe that the expressions for dimensions ($q,\,p$) can be obtained from those for ($p,\,q$) by changing the signs of all supertraces, $\str_j\to-\str_j$, as would be expected from the definition of supertrace. Moreover, as suggested by the fact that the coefficient of the identity always factors, we find that these matrix identities  can be written as the product of two matrix polynomials of degrees $p$ and $q$, respectively. This might be expected from the form of the reduced polynomial $P(x)=(a(x)+r)(d(x)+t)$,  but it is not an entirely obvious conclusion. The ($2,\,1$) identity, for example, is the product of the matrix polynomials $P_2, P_1$:
\bea
P_2&=&\left({\str_1}^2-{\str_2}\right) M^2-\frac{2}{3}
   \left({\str_1}^3-{\str_3}\right) M+\frac{1}{6} \left({\str_1}^4-4\,
   {\str_1}\, {\str_3}+3\, {\str_2}^2\right)\,I, \nn\\
 P_1&=&   \left({\str_1}^2-{\str_2}\right)M+\frac{1}{3} \left({\str_1}^3-3 \,{\str_1}\,
   {\str_2}+2\, {\str_3}\right)\,I. \nn
\eea

 The expressions for higher ($p,\,q$) values are exceedingly complex. 
However, for supermatrices with particular symmetries it is possible that all coefficients $a_k$ in \bref{superCH} have a common factor -- a polynomial in the supertraces --
leading to an identity of total degree lower than $2pq+p+q$. 
This happens for the $OSp(p|q)$ supermatrices $Z$. They are graded antisymmetric and 
satisfy $\Omega Z+Z^T \Omega=0$, where  $\Omega$ is the graded symmetric  $OSp$ metric and $Z^T$ is supertranspose of $Z$. More explicitly,  assuming covariant $Z$ indices,
\be
Z=\pmatrix{A&B\cr C&D},\quad
Z^T=\pmatrix{A^t&-C^t\cr -B^t&-D^t},\quad\Omega=\pmatrix{1&0\cr 0&J},\quad 
\ee 
\be
A+A^t=0,\quad D-D^t=0,\quad B=C^t,\quad  J+J^t=0.
\ee
Then the  $OSp(p|q)$  matrices $M=\Omega Z$ with mixed indices are graded antisymmetric and the  supertraces of all  odd powers of 
$M$ vanish.  Thus, for example, putting $\str_1=0=\str_3$ in the general $(2,\,2)$ case \bref{CH22}  we find an identity of  8th degree in the matrix elements of $OSp(2|2)$, after dividing out a common factor ${\str_2}^2$ :
 \bea
(2,\,2)_{OSp}&:& {\str_2}^2\,M^4-\str_2\, \str_4\, M^2 +\frac{1}{16} \left(-{\str_2}^4+ 4\,{\str_4}^2\right)I=0.
 \eea  
   
Trying to determine the $a_{k}$ polynomials for the general $(2,\,4)$ case requires massive amounts of memory, as it requires manipulating 22nd degree polynomials in 6 variables; on the other hand, if we impose the extra conditions  appropriate to
 the $OSp(2|4)$ case, we find  there is no unique solution for the  $a_{k}$ in the super Cayley-Hamilton equation \bref{superCH}.
However, the non-uniqueness is an overall factor of degree 8 in  \bref{superCH}, resulting in a unique identity of 14th degree in the matrix elements of $OSp(2|4)$: 
  \bea
(2,\,4)_{OSp}&:& \left({\str_2}^2 + 2\,{\str_4}\right)^2 M^6  \nn  \\&+&\frac{1}{6} \left({\str_2}^2 + 2\,{\str_4}\right) \left({\str_2}^3-6\,{\str_2}\,{\str_4} -16\,{\str_6}\right)M^4\nn \\ &-&\frac{1}{72}[{\str_2}^6+18\,{\str_2}^4{\str_4}-36\,{\str_2}^2{\str_4}^2 \nn  \\&+&64\,{\str_2}^3{\str_6}-72\,{\str_4}^3-128\,{\str_6}^2]M^2
   \\ &-&\frac{1}{144}[{\str_2}^7+6\,{\str_2}^5{\str_4}+12\,{\str_2}^3{\str_4}^2+72\,{\str_2}\,{\str_4}^3  \nn\\ &-& 8\,{\str_2}^4{\str_6} -96\,
  {\str_2}^2{\str_4}\,{\str_6}+96\,{\str_4}^2{\str_6}-128\,{\str_2}{\str_6}^2]\,I=0.\nn
  \eea

 When the $A,\,D$ submatrices have common eigenvalues,   the identities obtained above are still  meaningful, although the coefficients of the different powers of $M$ have zero bodies.  Thus these identities cannot be used for expressing higher powers of $M$ in terms of lower ones. 
  It seems likely (see the examples in  \cite{Morales94}) that for degenerate eigenvalues, identities with matrix powers higher than $p+q$ may be needed, but never more than $2pq+p+q$. 
 \vskip 3mm

  Having convinced ourselves of the validity of conjecture \bref{superCH}, we find generating functions for calculating the coefficients. We present  the results  without
proof since they were obtained by generalizing formulas obtained using  
diagonal supermatrices. We first write the characteristic equation \bref{superCH}
for $(p,q)$ supermatrices  as (assuming that the leading coefficient
$a_{2pq}$ has non zero body)
\be
\sum_{j=0}^{p+q}\,b^{(p,q)}_j\,M^{p+q-j}=0,\qquad b^{(p,q)}_0=1. 
\ee
The coefficients $b^{(p,q)}_j(S)$ are determined  by a generating function $F^{(p,q)}(S, t)$:
\be
F^{(p,q)}(S,t)=\sum_{j=0}^\infty b^{(p,q)}_j(S)\,t^j,\qquad 
b^{(p,q)}_j(S)=\frac1{j!}\left[\partial_t^j\,F^{(p,q)}(S,t)\right]_{t=0}.
\ee
The generating function for the case $q=0$  (ordinary matrix) is\footnote{ It is obtained most easily using the identity $det(e^Z)=e^{Tr Z}$, with $Z=\log(I-x A)$ and expanding the logarithm around $x=0$.}  \cite{genFunc}
\be
F^{(p,0)}(S,t)=G(S,t)\equiv e^{-\sum_{i=1}^\infty\frac{S_i\,t^i}{i}}.
\label{genp0}\ee
For $q>0$ we find that the generating function takes the form
\be
F^{(p,q)}(S,t)=\left(1-\sum_{k=1}^q\,\mu_k\,t^k\right)^2\,G(S,t),
\label{m1pq}\ee
where the $\mu_k$'s  are determined by the  linear system of equations
\be
 {\cal B}\,\pmatrix{\mu_1\cr \mu_2\cr ...\cr
\mu_q}
=\pmatrix{b_{p+1}\cr b_{p+2}\cr ...\cr
b_{p+q}},\qquad {\cal B}\equiv \pmatrix{b_p&b_{p-1}&...&b_{p-q+1} \cr
         b_{p+1}&b_{p}&...&b_{p-q+2} \cr ... \cr
b_{p+q-1}&b_{p+q-2}&...&b_{p} }. \label{m1qeq}
\ee
In this equation $b_j$ is an abbreviation for $b_j^{(p+q,0)}(S)$.
We have checked that the matrix ${\cal B}$ becomes singular  only when 
an eigenvalue of  submatrix $A$ coincides with one of $D$, 
so generating functions can be obtained only for the non degenerate case assumed in writing \bref{superCH}. We also assume that $q\leq p$. 
 The coefficients $b^{(p,q)}_j$ for  $q>p$ are obtained from the corresponding $b^{(q,p)}_j(S)$ by reversing the signs of all $S_j$'s  as noted above.

The  explicit form of  the generating function for $q=1$ is
\be
F^{(p,1)}(S,t)=(1-\mu_1\,t)^2\,G(S,t),
\qquad \mu_1=\frac{b^{(p+1,0)}_{p+1}(S)}{b^{(p+1,0)}_{p}(S)}
\label{m1p22}\ee
 and for $q=2$ is 
\be
F^{(p,2)}(S,t)=(1-\mu_1\,t-\mu_2\,t^2)^2\,
G(S,t),
\label{genp2}\ee
with
\be
 \mu_1=
   \frac{b^{(p+2,0)}_p b^{(p+2,0)}_{p+1}-b^{(p+2,0)}_{p-1} b^{(p+2,0)}_{p+2}}{(b^{(p+2,0)}_p)^2-b^{(p+2,0)}_{p-1}
   b^{(p+2,0)}_{p+1}},\quad 
\mu_2= \frac{b^{(p+2,0)}_pb^{(p+2,0)}_{p+2}-(b^{(p+2,0)}_{p+1})^2}
{(b^{(p+2,0)}_p)^2-b^{(p+2,0)}_{p-1} b^{(p+2,0)}_{p+1}}.\label{m1m2}
\ee
It should be pointed out that in equations \bref{m1pq} to \bref{m1m2} all $S_j$'s  that appear in the final expressions for the generating function $F^{(p,q)}(S,t)$ must be interpreted as supertraces
 $str(M^j)$ of the relevant $(p,\,q)$  supermatrix $M$. 
For example, the $\str_i$ in 
$  
b^{(4,0)}_3=\frac{1}{6}(-{\str_1}^3+3\,\str_1\,\str_2-2\,\str_3),
$ 
 which enters in both  $F^{(3,1)}(S,t)$ and $F^{(2,2)}(S,t)$, are different quantities in the two cases.
  
 These $b^{(p,q)}_j(S)$ coefficients, obtained from the corresponding generating functions, are rational functions of the supertraces. The polynomial coefficients 
$a_k$ in \bref{superCH} are obtained by multiplying   the corresponding $b^{(p,q)}_j$'s by $(\det{\cal B})^2$, 
 as the generating function is at most quadratic in the $\mu_k$'s. 
 The first coefficient $a_{2pq}=(\det{\cal B})^2$ in  \bref{superCH} is
indeed homogeneous of order $2pq$ in the matrix elements, because the $b^{(p,q)}_j$'s are homogeneous of order $j$.
Similarly, the coefficient $a_{2pq+1}$, depending on the term linear in $t$ in the generating function,  always has an overall factor of $\det{\cal B}$.

\vskip 3mm

 The results reported in this note present strong evidence for the validity of conjecture \bref{superCH}.  The generating functions obtained in \bref{m1pq} and \bref{m1qeq} are simple and can be used to derive general properties of the coefficients.  It would be most desirable if this conjecture  could be proved, and if  the expressions for the coefficients given above could be derived rigorously.

\subsection*{Addendum}

After this work was completed we became aware of similar results in the mathematical literature \cite{Kantor2002, Voronov2005}. However, the final matrix identities  obtained in these references, due to their derivation,  contain the supertrace of one higher power ($S_{p+q+1}$) than the dimension of the supermatrix. As a result these matrix identities, as given, cannot be factored into a product of two matrix polynomials of degrees $p$ and $q$. Of course, when $\str_{p+q+1}$ is replaced by its expression in terms of $\str_1,...,\str_{p+q}$,  the form of our identity \bref{superCH} (up to a factor $\det{\cal B}$) is obtained.

\end{document}